\documentclass[twocolumn,showpacs,preprintnumbers,amsmath,amssymb,prb]{revtex4}
\usepackage{graphicx}% Include figure files
\usepackage{dcolumn}% Align table columns on decimal point
\usepackage{bm}% bold math

\begin{document}

\preprint{Submitted to Phys. Rev. B}

\title{Electronic, dynamical, and thermal properties of ultra-incompressible superhard rhenium diboride: 
A combined first-principles and neutron scattering study }% Force line breaks with \\

\author{W. Zhou$^{1,2}$}
\author{H. Wu$^{1,3}$ }
\author{T. Yildirim$^{1,2}$}%
\affiliation{%
$^{1}$NIST Center for Neutron Research, National Institute of Standards and
Technology, Gaithersburg, Maryland 20899, USA\\
$^{2}$Department of Materials Science and Engineering, University of
Pennsylvania, Philadelphia, Pennsylvania 19104, USA \\
$^{3}$Department of Materials Science and Engineering, University of Maryland, College Park, Maryland 20742, USA}%

\date{\today}% It is always \today, today,

\begin{abstract}
Rhenium diboride is a recently recognized ultra-incompressible superhard material. 
Here we report the electronic (e), phonon (p), e-p coupling and thermal properties of 
ReB$_2$ from first-principles density-functional theory (DFT) calculations and neutron scattering measurements. 
Our calculated elastic constants ($c_{11}$ = 641 GPa, $c_{12}$ = 159 GPa, $c_{13}$ = 128 GPa, $c_{33}$ = 1037 GPa, and $c_{44}$ = 271 GPa), 
bulk modulus ($B$ $\approx$ 350 GPa) and hardness ($H$ $\approx$ 46 GPa) are in good agreement with the reported experimental data. 
The calculated phonon density of states (DOS) agrees very well with our neutron vibrational spectroscopy result. Electronic and phonon analysis 
indicates that the strong covalent B-B and Re-B bonding is the main reason for the super incompressibility and hardness 
of ReB$_2$. The thermal expansion coefficients, calculated within the quasi-harmonic 
approximation and measured by neutron powder diffraction, are found to be nearly isotropic 
in $a$ and $c$ directions and only slightly larger than that of diamond in terms of magnitude. 
The excellent agreement found between calculations and experimental measurements indicate 
that first-principles calculations capture the main interactions in this class of superhard 
materials, and thus can be used to search, predict, and design new materials with desired properties.
\end{abstract}

%Use showkeys class option if keyword
\maketitle
Hard materials are of great scientific interest due to their numerous 
technological applications. Unfortunately, almost all superhard materials 
(diamond, cubic BN etc.) are expensive because they either occur naturally 
in limited supplies or have to be made at high pressure synthetically. 
Therefore, intense research efforts have been carried out to design 
superhard materials\cite{Kaner:2005}. Recently, it was found that rhenium 
diboride can be synthesized at ambient pressure with potentially low cost, 
and the resulting ReB$_{2}$ crystal has super incompressibility along the 
$c$ axis, comparable to that of diamond, and high hardness, comparable to that 
of cubic BN.\cite{Chung:2007} The mechanical properties of ReB$_{2}$ were 
also correctly predicted by a recent theoretical work\cite{Hao:2006}. To 
more fully understand this unusual material, we have used a combined 
first-principles and neutron scattering study to further investigate the 
electronic, elastic, phonon and thermal properties of ReB$_{2}$.

Our calculations were performed within the plane-wave implementation of the 
generalized gradient approximation (GGA) to density-functional theory (DFT) 
in the PWscf package\cite{Baroni:1}. We used Vanderbilt-type ultrasoft 
potential with Perdew-Burke-Ernzerhof exchange correlation. A cutoff energy 
of 680 eV and a 16$\times $16$\times $6 $k$-point mesh (generated using the 
Monkhosrt-Pack scheme) were found to be enough for total energy to converge 
within 0.01 meV/atom. Spin-polarized calculations resulted in zero 
spontaneous spin polarization for the material investigated; thus here we 
focus on the results from spin-restricted calculations.

\begin{figure}
\includegraphics{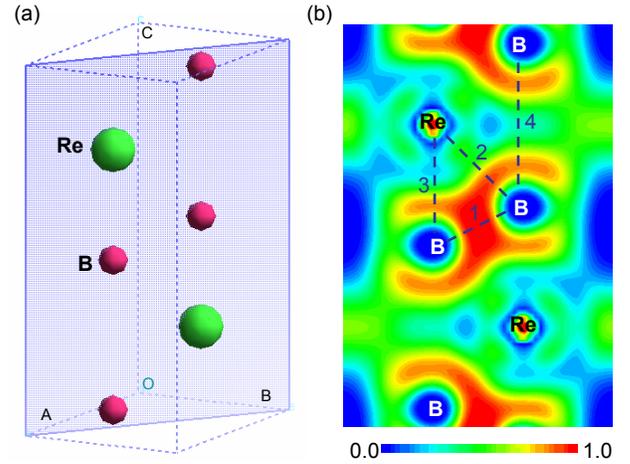}
\caption{
(Color online) (a) A unit cell of ReB$_2$. (b) The electron localization function of the (1 1 -2 0) 
crystal plane. The four types of chemical bonds are denoted using dashed lines. 
The strong directional, covalent B-B (labeled "1") and Re-B (labeled "2" and "3") bonding are apparent.
}
\label{fig1}
\end{figure}

Neutron scattering measurements were undertaken at the NIST Center for 
Neutron Research. The inelastic neutron scattering (INS) measurements were 
performed with the BT-4 filter-analyzer neutron spectrometer. The neutron 
powder diffraction was performed with the BT-1 high-resolution powder 
diffractometer. ReB$_{2}$ powder sample used in both experiments was 
synthesized using the method of direct heating of the elements (Re and 
$^{11}$B) in vacuum at 1273 K, as reported in Ref.2. The purpose of using 
$^{11}$B isotope is to avoid the large neutron absorption of normal 
boron\cite{Regular:1}.

ReB$_{2}$ has a simple hexagonal structure (space group $P6_{3}$\textit{/mmc}) with 
experimental $a$ = 2.900 {\AA} and $c$ = 7.478 {\AA}\cite{Placa:1962}, as shown 
in Fig. 1(a). The two Re atoms occupy the sites (1/3, 2/3, 1/4) and (2/3, 
1/3, 3/4) while the four boron atoms occupy the sites (2/3, 1/3, z), (1/3, 
2/3, 1/2+$z)$, (2/3, 1/3, 1/2-$z)$, (1/3, 2/3, 1-$z)$ with experimental $z$ = 0.048. We 
first optimized both the lattice parameters and the atomic positions of the 
ReB$_{2}$ structure. The relaxed parameters (without considering the zero 
point motion), $a$ = 2.9007 {\AA}, $c$ = 7.4777 {\AA}, and $z$ = 0.0478, all agree 
very well with the experimental values. We note that the crystal structure 
of ReB$_{2}$ is quite different from other metal diboride compounds, a 
majority of which assume MgB$_{2}$ type structure. In MgB$_{2}$, boron atoms 
are on the same plane, with a graphene 
structure\cite{Nagamatsu:2001}$^{,}$\cite{Yildirim:2001}, while they are 
buckled in ReB$_{2}$. In addition, Mg is located above the center of 
B-hexagon in MgB$_{2}$, while in ReB$_{2}$, Re is located right on top of 
the B atom. The structural configuration of ReB$_{2}$ is mainly due to the 
strong hybridization of Re-$d$ and B-$p$ orbitals, which we will discuss in detail 
below. 

\begin{table}[htbp]
\begin{center}
\caption{
The calculated bond population,  bond length, and real-space force constants 
for selected pairs of ions in a (1,1,-2,0) plane  in ReB$_2$. 
The positive (negative) bond population indicates bonding (antibonding) character. 
Note that the force constants are very large for nearest neighbor 
B-B and Re-B  pairs, indicating very strong local covalent bonding.}
\begin{tabular}{|l|l|c|c|c|}
\hline\hline
No.	& Bond	& Population &	Length (\AA) &	Force Constant (eV/\AA$^{2}$) \\ \hline
1 &	B-B	& 1.96	& 1.820 &	4.55 \\ 
2 &	Re-B    & 0.72  & 2.257	&      4.34 \\ 
3 &	Re-B	&-0.25	&2.227	&5.08 \\ 
4 &	B-B	&-0.14	&3.025	& 1.36 \\
\hline\hline
\end{tabular}
\label{tab1}
\end{center}
\end{table} 

To understand the high hardness of ReB$_{2}$, it is essential to look at its 
electronic structure. According to the calculated valence charge 
distribution, there exist large electron densities between two neighboring B 
atoms, and between the Re atoms and its neighboring B atoms, indicating 
strong directional, covalent B-B and Re-B bonding. This is as expected, 
since a material with large hardness must contain highly directional, short 
and strong 
bonds\cite{Cumberland:2005}$^{,}$\cite{Kaner:2005}$^{,}$\cite{Chung:2007}. 
To get more detailed information about the bonding nature, we performed 
electron localization function (ELF) analysis\cite{Savin:1997} and Mulliken 
bond population analysis\cite{Segall:1996}. In particular, the ELF plot is 
very useful in terms of distinguishing different bonding interactions. The 
value of ELF is in the range of 0 to 1 by definition. High ELF value 
corresponds to a low Pauli kinetic energy, as can be found in covalent bond. 
A value of ELF near 0.5 corresponds to delocalized electron density as found 
in metallic bonding. The ELF plot of the (1 1 -2 0) crystal plane of 
ReB$_{2}$ is shown in Fig. 1(b). The dominant feature is the rather strong 
B-B bond (labeled ``1''). The two Re-B bonds (labeled ``2'' and ``3'') 
clearly have different bonding characters. The B-B bond along the $c$ axis 
(labeled ``4'') has a large length, but ELF shows that its contribution to 
the overall bonding is not negligible. More quantitative bond population 
analysis (see Table I) confirms that the B-B bonding is very strong and 
highly covalent, with essentially a double bond nature. Interestingly, the 
Re and its two neighbored B atoms sitting right above and below it (along 
the $c$ axis) form anti-bonding, while Re forms strong covalent bonding with 
all other six B neighbors. In table I, the real-space force constants 
between these ions, obtained from the phonon calculation (discussed later), 
are also shown, which indicate strong B-B and Re-B bonding as well. The B-B 
and Re-B bonds, all together, form a covalently bonded three dimensional 
network.

\begin{figure}
\includegraphics{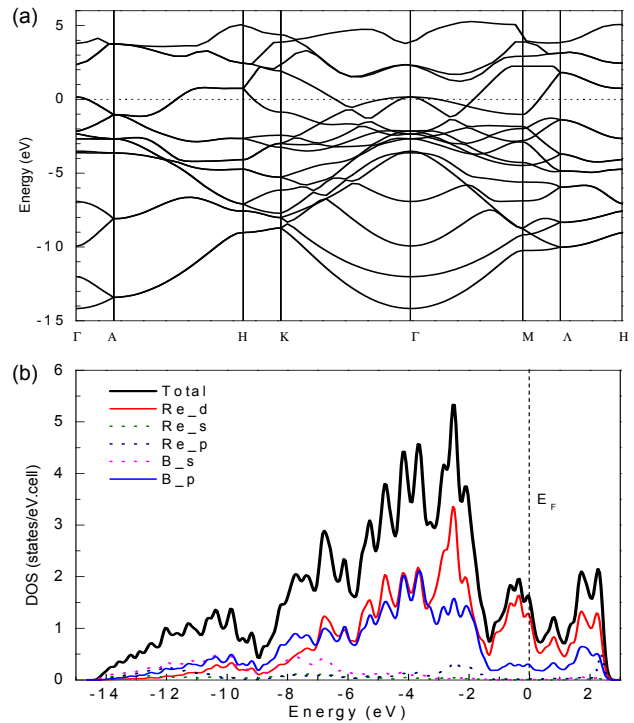}
\caption{
(Color online)
(a) The calculated electronic band structure of ReB$_2$, along high-symmetry directions 
in the Brillouin zone. $\Gamma$, A, H, K, M and $\Lambda$ represent k points (0 0 0), (0 0 0.5), 
(-1/3, 2/3, 0.5), (-1/3, 2/3, 0), (0, 0.5, 0) and (0, 0.5, 0.5) respectively. 
(b) The total and partial density of states (DOS) of ReB$_2$. The main feature of 
the electronic structure of ReB$_2$ is a strong hybridization of Re-$d$ and B-$p$ orbitals, 
which results in the strong covalent Re-B bond. 
}
\label{fig2}
\end{figure}

In Fig. 2(a), we show the electronic band structure of ReB$_{2}$. The total 
density of states (DOS) and the partial DOS projected onto atomic orbitals 
are shown in Fig. 2(b). Clearly, the material is metallic and the electronic 
structure of ReB$_{2}$ near the Fermi level is governed by a strong 
hybridization of Re-$d$ and B-$p$ orbitals, which results in the strong covalent 
bonding of B-B and Re-B. In terms of this bonding characteristic, ReB$_{2}$ 
is similar to the previously well studied 
OsB$_{2}$,\cite{Chen:2006}$^{,}$\cite{Hebbache:2006}$^{,}$\cite{Chiodo:2006}$^{,}$\cite{Gou:2006} 
although the latter has an orthorhombic structure. We also note that the 
density of states at the Fermi level, N(E$_{F})$ = 1.6 states/eV, is quite 
high and slightly larger than that of the 40 K superconducting 
MgB$_{2}$\cite{Yildirim:2001}, raising the question of possible 
superconductivity in ReB$_{2}$, which we will address later. 

\begin{figure}
\includegraphics{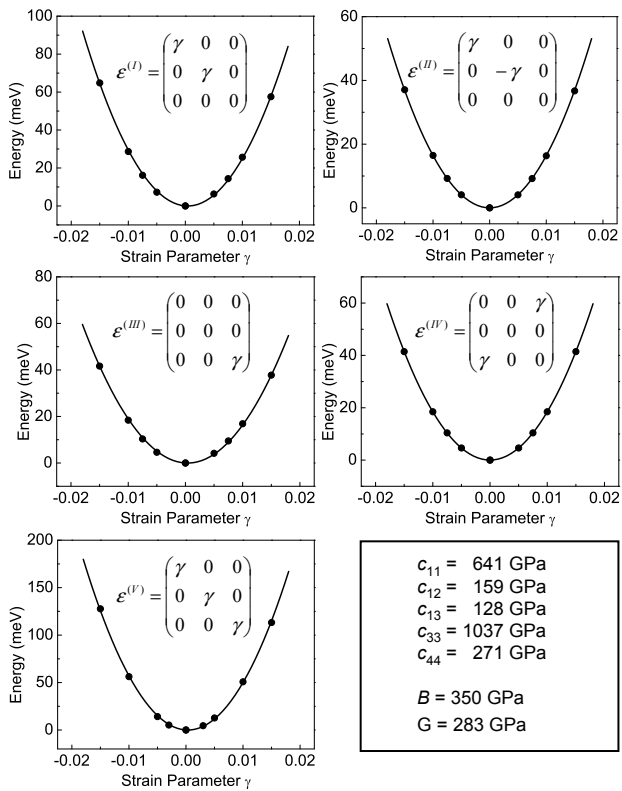}
\caption{
Least-square fit of the total energy data vs. strain parameters. 
The dots are from the first-principles calculations and the solid lines 
are the least-square fit. The deformation matrices for each distortion 
are also shown. The five hexagonal elastic constants of ReB$_2$, along 
with the bulk modulus and shear modulus are summarized in the right-bottom panel.
}
\label{fig3}
\end{figure}

We next discuss the elastic constants of the relaxed ReB$_{2}$ structure, 
which was calculated directly. For hexagonal crystal, there are five 
independent elastic coefficients: $c_{11}$, $c_{12}$, $c_{13}$, $c_{33}$ and 
$c_{44}$. Therefore, we applied five symmetry-independent 
strains\cite{Fast:1995}: ($\gamma $, $\gamma $, 0, 0, 0, 0), ($\gamma $, 
-$\gamma $, 0, 0, 0, 0), (0, 0, $\gamma $, 0, 0, 0), (0, 0, 0, 0, 2$\gamma 
$, 0), ($\gamma $, $\gamma $, $\gamma $, 0, 0, 0) to extract the five 
unknowns. We used $\gamma $ of $\pm $ 0.005, $\pm $ 0.0075, $\pm $ 0.01, 
$\pm $ 0.015. From the least-square fit of the total energy vs. strain 
data\cite{Page:2001} (Fig.3), we found that $c_{11}$ = 641 GPa, $c_{12}$ = 
159 GPa, $c_{13}$ = 128 GPa, $c_{33}$ = 1037 GPa and $c_{44}$ = 271 GPa. The 
quality of the data fit is excellent. By doubling the strain magnitudes, the 
fitted elastic constants only change slightly, suggesting that the 
anharmonicity is negligible in this structure. Note that our calculated 
elastic constant along $c$ axis, $c_{33}$ = 1037 GPa, is very large and indeed 
comparable to that of diamond ($\approx $ 1040 GPa, the largest known 
elastic coefficient\cite{Grimvall:1986}). This result conforms well with 
the reported experimental observation, namely, that the incompressibility 
along the $c$ axis is equal in magnitude to the linear incompressibility of 
diamond\cite{Chung:2007}. Also note the anisotropy in the 
compressibility of the two different lattice directions, i.e., $a$ axis is 
substantially more compressible than $c$ axis, which is also consistent with 
experimental observation. For a hexagonal crystal, the Voigt bulk modulus 
$B$ = 2($c_{11}+c_{12}$ + 2$c_{13}+c_{33}$/2) / 9, thus we obtain $B$ = 350 
GPa, in good agreement with the experimental value\cite{Chung:2007} of 
$\approx $ 360 GPa. The Voigt shear modulus $G$ = (7$c_{11}$ - 5$c_{12}$ - 
4$c_{13}$ + 2$c_{33}$ + 12$c_{44}$ ) / 9, from which we obtain $G$ = 283 GPa. 
Comparing our calculated elastic coefficients with Hao \textit{et. al.}'s computational 
results\cite{Hao:2006}, we found only a small difference ($<$ 5 
{\%}), which is likely due to the slight difference between different plane 
wave DFT codes.

Materials with high bulk and shear modulus are often hard materials. The 
hardness of a covalent or ionic compound can be directly calculated using 
the method recently proposed by \v{S}imùnek and 
Vack\'{a}ø\cite{imùnek:2006}, based on analyzing bond strengths and their 
densities. As discussed earlier, in the ReB$_{2}$ crystals, each Re atom 
forms eight bonds with B atoms, six of which have $d_{Re-B}$ = 2.257 {\AA} 
(bonding) while other two have slightly shorter $d_{Re-B}$ = 2.227 {\AA} 
(anti-bonding). In addition, each B atom forms three covalent bonds with its 
neighboring B atoms ($d_{B-B}$ = 1.820 {\AA}). The Re-Re interaction has a 
weak metallic nature with a nearest neighbor distance of 2.90 {\AA}, thus 
their contribution to the hardness is negligible. The reference 
energies\cite{imùnek:2006} (number of valence electron divided by the 
radius which makes the atom neutral) for Re and B were calculated to be 
4.878 and 3.09 respectively. Using equation (6) in Ref.19, we derived $H$ = 
46.0 GPa. Although the anisotropy of the hardness was not considered here, 
the calculated average value falls right within the range of the 
experimentally measured hardness ($\approx $ 30 to 56 
GPa)\cite{Chung:2007}.

Next, we discuss the phonon structure of ReB$_{2}$, which also reflects the 
mechanical properties. In addition, the lattice dynamics determine a wide 
range of other macroscopic behaviors such as thermal and transport 
properties, and the interaction with radiation (e.g., infra-red absorption, 
Raman scattering or inelastic neutron scattering). We performed the 
dynamical calculations on the optimized ReB$_{2}$ structure using the 
supercell method with finite difference.\cite{Yildirim:2000} A 2$\times 
$2$\times $1 supercell was used, and the full dynamical matrix was obtained 
from a total of 8 symmetry-independent atomic displacements (0.02 {\AA}). 
The unit cell of ReB$_{2}$ contains six atoms, which give rise to a total of 
18 phonon branches. The phonon modes at $\Gamma $ are classified as
\begin{eqnarray}
\Gamma (q=0) &=&  A_{1g}+ 2A_{2u} (IR) + 2B_{1g} + B_{2u} + E_{2u} \nonumber \\
&+&2 E_{2g}(R) + 2 E_{1u} (IR) + E_{1g} (R), \nonumber
\end{eqnarray}
where $R$ and \textit{IR} correspond to Raman- and infrared-active, respectively. The 
crystal symmetry implies six Raman- and six \textit{IR}-active modes. In Table II, we 
list the calculated energies at $\Gamma $ and corresponding mode characters.

The computed phonon dispersion curves and phonon density of states are shown 
in Fig. 4(a). The acoustic branches have steep slopes, indicating large 
elastic coefficients. Using the low energy part of the acoustic branches 
(i.e., ``elastic limit''), we can estimate the sound velocity ($V)$ and thus 
also elastic constants. For example, for a hexagonal crystal, $c_{33}$ = 
\textit{$\rho $V}(longitudinal [0001])$^{2}$ and $c_{44}$ = \textit{$\rho $V}(transverse [0001])$^{2}$, where 
\textit{$\rho $} is the mass density. We derived $c_{33}$ = 948 GPa and $c_{44}$ = 278 GPa. 
The slope determination close to $q$ = 0 certainly has some error bar; 
nevertheless, the above estimated values are still in reasonable agreement 
with the more precise numbers calculated earlier by directly applying 
strains.

The optical phonon branches are clearly divided into two groups. The three 
lowest energy optical modes ($<$ 30 meV) are dominated by Re motion while 
the high energy optical modes ($>$ 50 meV) are dominated by the lighter B 
atom displacement. Note that for similar types of atom motion, we found that 
those along $c$ axis always have higher energies than the one within the basal 
$a-b$ plane. This is due to the covalent Re-B bonding along the $c$ axis. For Boron 
motion within the plane, those involving B-B bond stretching (i.e., the 
motion of the bonded B-B pair being out-of-phase) are of higher energy than 
the in-phase motion, due to the strong B-B covalent bonding. Hence, the 
phonon structure is totally consistent with our ealier electronic structure 
analysis.

\begin{table}[htbp]
\begin{center}
\caption{
The calculated phonon energies E and the characteristics of the optical modes of 
ReB$_2$ at $\Gamma$-point. Note that in the table, "in-phase" and "out-of-phase" refer 
to the movement of the bonded B-B pair in the unit cell. 
($IR$) and ($R$) indicates the $IR$ and Raman active modes, respectively. The last column indicates 
the electron-phonon coupling for each modes at  $\Gamma$-point, 
which are quite small compared to those of MgB$_2$.}
\begin{tabular}{|c|c|l|c|}
\hline\hline
Species &	E (meV)	& Dominant character/type	&	$\lambda_{el-ph}$ \\ \hline
2$E_{2g}$	&18.6	&Re in $a$-$b$ plane	&	0 \\
$B_{1g}$	&28.5	&Re along $c$		&0.007 \\
2$E_{2u}$	&50.1	&B in $a$-$b$ plane, in-phase &		0 \\
2$E_{1u}$	&59.6	&B in $a$-$b$ plane, in-phase	($IR$) &	0.008 \\
$A_{2u}$	&78.1	&B along $c$, in-phase	($IR$)  &	0 \\
2$E_{1g}$	&85.2	&B in $a$-$b$ plane, out-of-phase	($R$)	& 0.002  \\
$B_{2u}$	&87.6	&B along $c$, out-of-phase		 & 0.025  \\
2$E_{2g}$	&90.4	&B in $a$-$b$ plane, out-of-phase	($R$) &	0.016  \\
$B_{1g}$	&91.1	&B along $c$, out-of-phase		& 0.021 \\
$A_{1g}$	&97.7	&B along $c$, out-of-phase		& 0.018  \\
\hline\hline
\end{tabular}
\label{tab2}
\end{center}
\end{table} 

\begin{figure}
\includegraphics{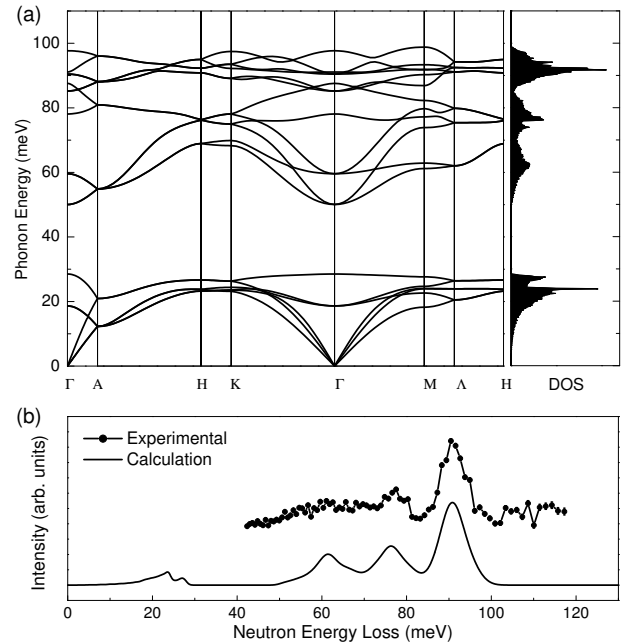}
\caption{
 (a) Calculated phonon dispersion curves along high-symmetry directions in the Brillouin 
zone for ReB$_2$ and the phonon density of states. (b) Inelastic neutron scattering 
spectrum from ReB$_2$ (at 4 K) along with the calculated spectrum. The agreement is excellent.
}
\label{fig4}
\end{figure}

The calculated phonon structure is further validated by the neutron 
spectroscopy measurement. The measured INS spectrum is essentially the 
phonon DOS weighted by the neutron scattering cross sections of the 
elements. As shown in Fig. 4(b), the agreement between the experimental data 
and the calculated spectrum based on the incoherent 
approximation\cite{Squires:1996} is excellent.

It is interesting to note that the phonon structure of ReB$_{2}$ is quite 
similar to that of MgB$_{2}$, a high-$T_{c}$ material\cite{Yildirim:2001}. 
With a MgB$_{2}$-like phonon spectrum and a slightly higher N(E$_{F})$ than 
that of MgB$_{2}$ as mentioned earlier, one may wonder if ReB$_{2}$ can 
exhibit superconductivity at a ``high'' temperature. Experimentally, in the 
Re-B system, Re$_{3}$B and Re$_{7}$B$_{3}$ were found to have a $T_{c}$ of 
4.8 K and 3.3 K, respectively.\cite{Kawano:2003} We are not aware of any 
experimental measurement of $T_{c}$ in ReB$_{2}$. We thus calculated the 
electron-phonon coupling parameters at $\Gamma $ for ReB$_{2}$ and tried to 
estimate $T_{c}$. The results are shown in Table II. The coupling mainly 
takes place for B phonons along $c$ axis. Compared to MgB$_{2}$, the 
electron-phonon coupling in ReB$_{2}$ is much weaker. Using the McMillian 
expression\cite{McMillan:1968} and only considering the zone center phonon, 
$T_{c}$ of ReB$_{2}$ is estimated to be less than 1 K. It would be 
interesting to confirm this estimate experimentally.

Finally, we discuss the thermal expansion of the ReB$_{2}$ structure, a 
property important for practical applications. The temperature dependence of 
the lattice parameters $a$ and $c$ are calculated within the quasi-harmonic 
approximation. The results were obtained by minimizing the Helmholtz free 
energy
\begin{eqnarray}
F(a,c,T) = V(a,c) &+& \sum_{j}\sum_{{\bf q}} \biggl \{ \frac{1}{2}\hbar \omega_{j}({\bf q}) \nonumber \\
&+&kT ln(1-e^{- \hbar \omega_{j}({\bf q})/kT} )\biggr \}, \nonumber
\end{eqnarray}
where the first term is the ground state energy and the second term is 
obtained by summing the phonon modes over the wavevectors in the Brillouin 
zone. In the quasi-harmonic approximation, the effect of the anharmonicity 
in the lattice energy is treated by allowing the phonon frequencies to 
depend on the lattice parameters. Hence, for a given temperature $T$, we first 
take $a$ and $c$, minimize the atomic positions, and then calculate the phonon 
spectrum, and the free energy. Repeating this for other values of $a$ and $c$, we 
find the optimum values of the lattice parameters that minimize the free 
energy at a given temperature. In this way, one obtains the temperature 
dependence of $a$ and $c$.

\begin{figure}
\includegraphics{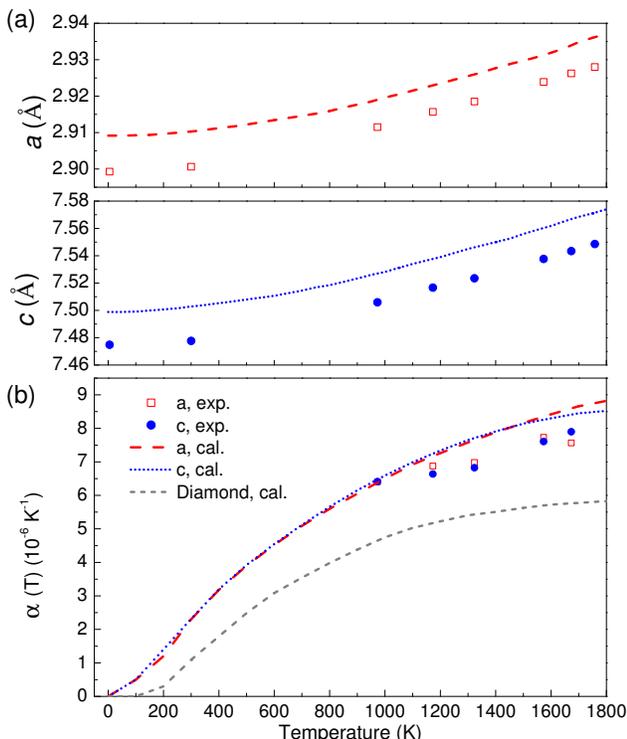}
\caption{
(Color online)
 (a) The calculated (lines) and experimental (dots) temperature dependence of 
the structural parameters $a$ and $c$ of ReB$_2$. (b) The calculated (lines) 
and experimental (dots) linear thermal expansion coefficient of ReB$_2$. 
For comparison purpose, the calculated linear thermal expansion coefficient 
of diamond is also shown. The thermal expansion of ReB$_2$ is nearly isotropic 
and the magnitude is in the same order as that of diamond. 
}
\label{fig5}
\end{figure}

The calculated temperature dependence is shown in Fig. 5(a), along with the 
experimental lattice constants measured by neutron powder diffraction. 
Although the DFT-GGA calculation predicts slightly larger lattice constants 
than the measured values, the overall agreement on the lattice expansion is 
reasonably good. We found that the linear thermal expansion coefficients 
along $a$ and $c$ axes are both very small [see Fig. 5(b)], $\approx $ 6.5$\times 
$10$^{-6}$ K$^{-1}$ at 1000 K, slightly larger than that of diamond 
($\approx $ 4.5$\times $10$^{-6}$ K$^{-1}$ at 1000 K). Additionally, $a$ and 
$c$ expansion coefficient curves nearly overlap with each other, suggesting 
that the material is highly isotropic in terms of thermal expansion, which 
is preferred in most applications. The excellent combination of hardness, 
incompressibility and small thermal expansion coefficient indicates that 
ReB$_{2}$ has great potential for many applications, such as material for 
cutting tools and abrasion resistance coating.

In summary, we have applied density-functional theory and neutron scattering 
techniques to elucidate the electronic, elastic, phonon, and thermal 
properties of ReB$_{2}$. Our calculated elastic constants, bulk modulus and 
hardness are in very good agreement with the experimental data. Our 
electronic and phonon results confirmed that the strong covalent B-B bonding 
and Re-B bonding play a critical role in the incompressibility and hardness 
of ReB$_{2}$. Our calculations indicate that ReB$_{2}$ has a very similar 
phonon spectrum to MgB$_{2}$ with a comparable N(E$_{F})$. However, we found 
a very small electron-phonon coupling, which suggests a very modest 
superconducting temperature. The thermal expansion coefficient is found only 
slightly larger than that of diamond. The combined excellent mechanical and 
thermal properties suggest great potential for ReB$_{2}$ to be used as a 
cutting or coating material. The excellent agreement found between DFT 
calculations and experimental measurements indicate that first-principles 
calculations are able to capture the main interactions in this class of 
superhard materials, and thus can be used to search, predict, and design 
other materials with properties (hardness etc.) better than diamond.

The authors thank T. J. Udovic for technical help in the INS data 
collection.


\begin{thebibliography}{8}
\bibitem{Regular:1} Regular Boron contains $\approx $ 20{\%} $^{10}$B and $\approx $ 80{\%} $^{11}$B. $^{10}$B isotope has a very large neutron absorption cross section.
\end{thebibliography}
\end{document}